\documentstyle[prl,aps,multicol,psfig]{revtex}

\renewcommand{\narrowtext}{\begin{multicols}{2} \global\columnwidth20.5pc}
\def\al{\alpha}

\def\de{\delta}

\def\ve{\varepsilon}

\def\et{\eta}
\def\th{\theta}

\def\la{\lambda}

\def\rh{\rho}

\def\si{\sigma}

\def\ta{\tau}

\def\ph{\phi}

\def\La{\Lambda}

\def\Om{\Omega}

\def\cl{{\cal L}}

\def\fr#1#2{{{#1} \over {#2}}}

\def\frac#1#2{{\textstyle{{#1}\over {#2}}}}

\def\lsim{\mathrel{\rlap{\lower4pt\hbox{\hskip1pt$\sim$}}
    \raise1pt\hbox{$<$}}}
\def\gsim{\mathrel{\rlap{\lower4pt\hbox{\hskip1pt$\sim$}}
    \raise1pt\hbox{$>$}}}
\def\sqr#1#2{{\vcenter{\vbox{\hrule height.#2pt
         \hbox{\vrule width.#2pt height#1pt \kern#1pt
         \vrule width.#2pt}
         \hrule height.#2pt}}}}

\def\prt{\partial}
\def\lrpartial{\raise 1pt\hbox{$\stackrel\leftrightarrow\partial$}}

\def\etal{{\it et al.}}

\newcommand{\beq}{\begin{equation}}
\newcommand{\eeq}{\end{equation}}
\newcommand{\bea}{\begin{eqnarray}}
\newcommand{\eea}{\end{eqnarray}}
\newcommand{\rf}[1]{(\ref{#1})}

\def\sech{\mathop{\rm sech}\nolimits}

\begin{document}

\title{Spacetime-varying couplings and Lorentz violation} 

\author{V.\ Alan Kosteleck\'y,$^a$
Ralf Lehnert,$^b$ and Malcolm J.\ Perry$^c$}

\address{$^a$Physics Department, Indiana University,
Bloomington, IN 47405, U.S.A.}
\address{$^b$Physics Department, Universidade do Algarve,
8000 Faro, Portugal}
\address{$^c$D.A.M.T.P., Cambridge University,
Wilberforce Road, Cambridge CB3 0WA, England}  

\date{IUHET 452, November 2002}

\maketitle

\begin{abstract}
Spacetime-varying coupling constants
can be associated with violations of local Lorentz invariance
and CPT symmetry.
An analytical supergravity cosmology 
with time-varying fine-structure constant 
provides an explicit example.
Estimates are made for some experimental constraints.
\end{abstract}


\narrowtext
Since Dirac's large-number hypothesis 
\cite{lnh},
spacetime-varying couplings have remained the subject
of various theoretical and experimental studies.
Such couplings are natural in many unified theories 
\cite{theo},
and current claims of observational evidence
for a time-varying electromagnetic coupling \cite{webb}
have sparked a revival of this idea \cite{jp}.

In this work,
we investigate the role of Lorentz symmetry
in the subject,
showing that spacetime-varying couplings
can be associated with Lorentz and CPT violation
\cite{cpt01}.
This result is intuitively reasonable
because translation invariance is broken 
in a theory with spacetime-varying couplings,
while translations and Lorentz transformations
are intertwined in the Poincar\'e group.
The vacuum then behaves as a spacetime-varying medium
so Lorentz isotropy can be lost in local inertial frames.

As an illustration,
consider a spacetime-varying coupling $\xi$
associated with a term containing derivatives
in a lagrangian $\cl$.
A simple example involving a scalar $\ph$ is a term 
$\cl\supset\xi\prt_{\mu}\ph\prt^{\mu}\ph$,
which implies 
$\cl\supset-\ph(\prt_\mu\xi)\prt^{\mu}\ph$
upon an integration by parts.
If $\xi$ varies smoothly, 
$\prt_\mu\xi$ has a piece 
that behaves in a local inertial frame
as a coefficient $k_\mu$ for Lorentz and CPT violation.
More generally,
non-scalar fields can play a role,
and the effects can arise through  
subsidiary conditions involving coefficients like $k_\mu$ 
appearing in the equations of motion.

All possible Lorentz-violating lagrangian terms 
are given by the Lorentz- and CPT-violating standard-model extension 
\cite{ck},
and many have been bounded experimentally
in precision experiments 
with hadrons
\cite{hadronexpt,hadronth},
protons and neutrons
\cite{pn},
electrons
\cite{eexpt,eexpt2},
photons
\cite{cfj,km},
and muons
\cite{muons}.
The theory contains all observer Lorentz scalars 
formed by combining operators 
and coefficients having Lorentz indices.
Terms of this type arise, for example,  
from spontaneous Lorentz violation
\cite{kps} 
and in realistic noncommutative field theories
\cite{ncqed}.
The presence of translation violations 
induced by spacetime-varying couplings 
complicates theoretical and experimental analyses.
Here, we focus on showing that spacetime-varying couplings
and apparent Lorentz violation can arise naturally,
even when the dynamics of the underlying theory 
is Lorentz invariant and involves only constant couplings.

Our analysis is performed in the context of
$N=4$ supergravity in four dimensions.
Although this model is unrealistic in detail,
it is a limit of the $N=1$ supergravity in 11 spacetime dimensions
and hence is a limit of M theory.
It can therefore shed light on generic features
to be expected in a fundamental theory.
We show that smoothly varying couplings 
can naturally be obtained from a simple cosmological solution.
In particular,
in electrodynamics the fine-structure constant $\al=e^2/4\pi$
and the $\th$ angle acquire related spacetime dependences,
driving the Lorentz violation.

The spectrum of the $N=4$ supergravity in four spacetime dimensions 
consists of the graviton,
represented by the metric $g_{\mu\nu}$,
four gravitinos,
six abelian graviphotons $A_\mu^{jk}$,
four fermions,
and a complex scalar $Z$
that contains an axion and a dilaton.
The Latin indices $j,k,\ldots$
denote vector indices in the SO(4) internal symmetry,
and the graviphotons lie in the adjoint representation.
The bosonic part $\cl$ of the lagrangian can be written 
\cite{cj}
\bea
\cl
&=& 
-\frac 1 2 \sqrt{g} R
-\frac 1 4 \sqrt{g} M_{jklm} F_{\mu\nu}^{jk} F^{lm\mu\nu}
\nonumber\\
&&
-\frac 1 8 \sqrt{g} N_{jklm}\ve^{\mu\nu\rh\si} 
F_{\mu\nu}^{jk} F^{lm}_{\rh\si}
+ \sqrt{g} \fr
{\prt_\mu Z \prt^\mu\overline{Z}}
{(1 - Z\overline{Z})^2},
\label{lag}
\eea
where Planck units are adopted.
The generalized electromagnetic coupling constant $M_{jklm}$ 
and the $\th$-term coupling $N_{jklm}$
are both real and determined by the complex scalar $Z$
according to
\beq
M_{jklm} + i N_{jklm} =
\fr{
\de_{[j|l|}\de_{k]m} 
(1-Z^2)
- i\ve_{jklm}Z }
{(1+Z^2)} .
\label{mplusn}
\eeq
For present purposes,
it is convenient to apply the 
Cayley map $W = -i(Z-1)/(Z+1)$ taking the unit disk 
into the upper half plane.
Writing $W=A+iB$,
the scalar kinetic term 
becomes
$\cl_{\rm b}=
\sqrt{g} (\prt_\mu A\prt^\mu A + \prt_\mu B\prt^\mu B)/4B^2$,
and $M$ and $N$ undergo corresponding transformations.
Then,
$B$ can be identified with the string-theory dilaton. 

We consider the case in which only one graviphoton,
$F^{12}_{\mu\nu}\equiv F_{\mu\nu}$,
is excited.
The bosonic lagrangian then becomes
\bea
\cl
&=& 
-\frac 1 2 \sqrt{g} R 
-\frac 1 4 \sqrt{g} M F_{\mu\nu} F^{\mu\nu}
-\frac 1 4 \sqrt{g} N F_{\mu\nu} \tilde{F}^{\mu\nu}
\nonumber\\
&& 
\qquad
+\sqrt{g} ({\prt_\mu A\prt^\mu A + \prt_\mu B\prt^\mu B})/{4B^2},
\label{lag2}
\eea
with $\tilde{F}^{\mu\nu}=\ve^{\mu\nu\rh\si}F_{\rh\si}/2$ and
\bea
M &=& \fr
{B (A^2 + B^2 + 1)}
{(1+A^2 + B^2)^2 - 4 A^2},
\nonumber\\
N &=& \fr
{A (A^2 + B^2 - 1)}
{(1+A^2 + B^2)^2 - 4 A^2}.
\label{N}
\eea

Consider a cosmology in this theory 
involving a flat ($k=0$) Friedmann-Robertson-Walker (FRW) model.
The line element for the associated spacetime is 
\beq
ds^2 = dt^2 - a^2(t) (dx^2 + dy^2 + dz^2) ,
\label{frw}
\eeq
where $t$ is the comoving time
and $a(t)$ is the cosmological scale factor.
The usual assumptions of homogeneity and isotropy imply that 
$A$ and $B$ are also functions only of $t$.
Solving the Einstein equations
with just the scalar field as a source of energy and momentum
yields $a(t) \sim t^{1/3}$,
which is an expansion rate far slower than seen in our Universe.
A standard approach to obtain a more realistic theory
adds an energy-momentum tensor 
$T_{\mu\nu} = \rh u_\mu u_\nu$
describing galaxies and other matter,
where $u^\mu$ is a unit timelike vector
orthogonal to spatial surfaces
and $\rh(t)$ is the energy density of the matter.
In our supergravity model,
an energy-momentum tensor of this form 
arises from the fermionic sector
because the fermion kinetic terms 
are uncoupled from the scalar field $W$,
and so $T_{\mu\nu}$ is independent of $W$.

Ignoring the graviphoton for the moment,
the Einstein equations for the supergravity cosmology
in the presence of the fermion matter are
\bea
G_{\mu\nu} &=& T_{\mu\nu}
+ \fr 1 {2B^2}
( \prt_\mu A\prt_\nu A + \prt_\mu B\prt_\nu B)
\nonumber \\ && 
\qquad
- \fr 1 {4B^2} g_{\mu\nu}
( \prt_\la A\prt^\la A + \prt_\la B\prt^\la B ).
\label{einst}
\eea
For the $k=0$ FRW model,
this expression contains only two independent equations:
\beq
- 3 \fr {\ddot a}{a} =
\frac 1 2 \rh + \fr 1 {2B^2} (\dot A^2 + \dot B^2),
\quad
\fr {\ddot a}{a} + 2 \fr {\dot a^2 }{a^2} =
\frac 1 2 \rh ,
\label{eq2}
\eeq
where a dot indicates a time derivative.
The system is also governed by
the equations of motion for $A$ and $B$:
\beq
\fr d {dt} 
\left(
\fr {a^3 \dot A}{B^2}\right) =0,
\quad
\fr d {dt} 
\left(
\fr {a^3 \dot B}{B^2} 
\right)
+ \fr {a^3 }{B^3} (\dot A^2 + \dot B^2) =0 . 
\label{eq4}
\eeq
The final equation determining the time evolution,
${d(\rh a^3)}/{dt} = 0$,
follows from conservation of energy.

It turns out 
these five equations can be integrated analytically.
Suppose that at the present time $t_n$
the Universe has matter density $\rh_n$
and scale size $a_n = a(t_n)$.
Energy conservation yields
$\rh(t) = c_n/a^3(t)$,
where $c_n = \rh_n a_n^3$.
Integration of one Einstein equation then gives
\beq
a(t) = 
\left(
\frac 3 4 c_n (t+t_0)^2 - c_1
\right)^{1/3} .
\label{at}
\eeq
Here,
$c_1$ is an integration constant describing the 
amount of energy in the scalar fields.
Also, 
$t_0$ is another integration constant,
chosen here as $t_0 = \sqrt{4c_1/3c_n}$
to fix the time origin $t=0$ 
at the moment of the initial singularity when $a(t) = 0$.
Note that for $t\gg t_0$ we find $a(t) \sim t^{2/3}$,
as expected for a $k=0$ matter-dominated Universe.

The equation of motion for $A$ 
can be integrated once to give
$\dot A = c_2 B^2/a^3$,
where $c_2$ is an integration constant.
The remaining equations can be solved
to yield a functional form for $A$ and $B$
in terms of a parameter time $\ta$.
This leaves two equations,
related through the Bianchi identities.
After some algebra, 
we find
\beq
A = \pm\la \tanh (\fr 1 \ta + c_3) + A_0,
\quad
B = \la \sech (\fr 1 \ta + c_3) ,
\label{be}
\eeq
where $\la\equiv \mp 4 c_1/\sqrt{3} c_2 t_0$,
and $c_3$, $A_0$ are integration constants.
The cosmological time $t$ is given in terms
of the parametric time $\ta$ by
$t = t_0 [\coth({\sqrt{3}}/{4\ta}) - 1]$,
so $t=0$ when $\ta = 0$
and $t$ increases when $\ta$ increases.
In what follows,
it suffices to adopt the simplifying choice $c_3=0$.
At late times $t\gg t_0$,
we then find $\ta \approx \sqrt{3} t/4t_0$,
$A \approx \pm 4 \la t_0/\sqrt{3} t + A_0$,
and $B \approx \la (1 - 8t_0^2/3t^2)$.
This means both $A$ and $B$ tend to constant values
at late times on a timescale set by $t_0$.
The supergravity cosmology 
therefore fixes the value of the string-theory dilaton,
despite the absence of a dilaton potential.

We next consider excitations of $F_{\mu\nu}$
in the axion-dilaton background \rf{be}.
For the moment,
we restrict attention to localized excitations
in spacetime regions that are small on a cosmological scale.
This corresponds to most experimental situations,
and it is therefore appropriate to work in a local inertial frame.

Allowing for a nontrivial $\th$ angle,
the conventional electrodynamics lagrangian in a local
inertial frame can be taken as 
\beq
\cl_{\rm em} = 
-\fr{1}{4 e^2} F_{\mu\nu}F^{\mu\nu}
- \fr{\th}{16\pi^2} F_{\mu\nu} \tilde{F}^{\mu\nu}.
\label{em}
\eeq
In the supergravity cosmology,
we can identify
$e^2 \equiv 1/M$, $\th \equiv 4\pi^2 N$.
Since $M$, $N$
are functions of the background fields $A$, $B$,
it follows that $e$, $\th$ acquire spacetime dependence
in an arbitrary local inertial frame.

The equations of motion 
in the presence of charged matter
described by a 4-current $j^{\nu}$ are
\beq
\fr{1}{e^2}\partial_{\mu}F^{\mu\nu}
-\fr{2}{e^3}(\partial_{\mu}e)F^{\mu\nu}
+\fr{1}{4\pi^2}(\partial_{\mu}\th)\tilde{F}^{\mu\nu}=j^{\nu} .
\label{Feom}
\eeq
In a trivial background,
the last two terms on the left-hand side of this equation
would vanish and the usual Maxwell equations would emerge.
Here, however,
the extra two terms lead to apparent Lorentz-violating effects
despite being coordinate invariant.
On small cosmological scales,
$\prt_\mu M$ and $\prt_\mu N$
are approximately constant,
and they therefore select a preferred direction 
in the local inertial frame.
This means that particle Lorentz symmetry,
as defined in the first paper of Ref.\ \cite{ck},
is broken.

Note that the expansion in a textbook FRW cosmology 
without scalar couplings lacks this violation 
because a local Lorentz-symmetric inertial frame always exists,
whereas in the present case the variation of $M$ and $N$
implies particle Lorentz violation in any local inertial frame.
Indeed,
the above cosmology-induced Lorentz violation 
is independent of the details of the
$N=4$ supergravity model.
Any similarly implemented smooth spacetime variation 
of the electromagnetic couplings
on cosmological scales leads to such effects.
This suggests particle Lorentz violation
could be a common feature of models
with spacetime-dependent couplings.

In the local inertial frame,
we can write
\beq
\cl_{\rm em}^{\prime} = -\fr{1}{4 e^2}  F_{\mu\nu} F^{\mu\nu}
+\fr{1}{8\pi^2}(\prt_\mu\th) A_\nu \tilde{F}^{\mu\nu} .
\label{prlagr}
\eeq
A nonzero constant contribution from $\prt_\mu \th$ 
demonstrates explicitly the violations of particle Lorentz invariance
and CPT symmetry.
To facilitate contact with the conventional notation
in the Lorentz-violating standard-model extension,
we can identify $(k_{AF})_\mu \equiv e^2 \prt_\mu\th/8\pi^2$.
In our supergravity model,
$(k_{AF})_\mu$ is timelike.

The special case of constant $e$ and constant 
$(k_{AF})_\mu$
has been discussed extensively in the literature 
\cite{cfj,ck,jkk}.
Under these conditions,
the lagrangian \rf{prlagr}
is invariant under spacetime translations,
but the associated conserved energy
fails to be positive definite 
and so leads to instabilities.
It is natural to ask how this difficulty is circumvented
in the present model,
which arises from a positive-definite supergravity theory 
\cite{fn1}.

A key difference is that,
instead of being nondynamical and constant,
$(k_{AF})_\mu$ depends in the present model
on the dynamical degrees of freedom $A$, $B$. 
Excitations with $F_{\mu\nu}\neq 0$
therefore cause perturbations $\de A$, $\de B$
away from the cosmological solutions \rf{be},
so that $A\rightarrow A+\de A$
and $B\rightarrow B+\de B$.
It follows that $\th\rightarrow \th+\de\th$
and that the energy-momentum tensor $(T^{\rm b})^{\mu\nu}$ 
of the background
receives an additional contribution,
$(T^{\rm b})^{\mu\nu}\rightarrow (T^{\rm b}_F)^{\mu\nu}=
(T^{\rm b})^{\mu\nu}+\de (T^{\rm b})^{\mu\nu}$.
This contribution can compensate 
for negative-energy ones from the $(k_{AF})_\mu$ term.

The compensation mechanism can be 
illustrated explicitly at the classical level
in the lagrangian
$\cl=\cl_{\rm em}^{\prime}+\cl_{\rm b}$
\cite{fn2}.
The relevant feature for present purposes
is the $A$- and $B$-dependence of $\th$,
so for simplicity $e$ can be taken as constant.
We begin
by splitting the total conserved energy-momentum tensor
into two pieces,
$(T^{\rm t}_F)^{\mu\nu}=(T^{\rm em})^{\mu\nu}+(T_F^{\rm b})^{\mu\nu}$,
where
\bea
(T^{\rm em})^{\mu\nu} & = & 
\fr{\prt \cl}{\prt(\prt_{\mu}A^{\la})}\,\prt^{\nu}\! A^{\la}
-\et^{\mu\nu}\cl_{\rm em}^{\prime} ,\nonumber\\
(T_F^{\rm b})^{\mu\nu} & = & 
\fr{\prt \cl}{\prt(\prt_{\mu}A)}\,\prt^{\nu}\! A
+\fr{\prt \cl}{\prt(\prt_{\mu}B)}\,\prt^{\nu}\! B
-\et^{\mu\nu}\cl_{\rm b} .
\label{split}
\eea
Explicitly, we find
\bea
(T^{\rm em})^{\mu\nu} & = & 
\fr{1}{e^2}F^{\mu}_{\;\;\la}F^{\la\nu}
+ \fr{1}{4e^2}\et^{\mu\nu}F^{\rh\si}F_{\rh\si}
\nonumber\\
&&\qquad
+\fr{1}{8\pi^2}(\prt^\nu\th)A_{\la}\tilde{F}^{\la\mu} .
\label{emex}
\eea
Negative-energy contributions can arise only from the last term.
Similarly, we obtain 
\bea
(T_F^{\rm b})^{\mu\nu} & = & 
\fr {\prt^\mu A\prt^\nu A}{2B^2}
-\fr {\et^{\mu\nu}} {4B^2}
( \prt_\la A\prt^\la A + \prt_\la B\prt^\la B )
\nonumber\\
&& \qquad +\fr{\prt^\mu B\prt^\nu B}{2B^2} 
-\fr{1}{8\pi^2}(\prt^\nu\th)A_{\la}\tilde{F}^{\la\mu},
\label{bex}
\eea
where again only the last term can lead to negative-energy contributions.
Combining the two equations shows that 
the total conserved energy is positive definite,
even when a nonzero $(k_{AF})_\mu$ is generated.
The apparent paradox arises only because
the two pieces $(T_F^{\rm em})^{\mu\nu}$ and $(T_F^{\rm b})^{\mu\nu}$,
each with positivity difficulties,
are separately conserved when $\prt^\nu\th$ is constant
\cite{fn3}.

Another interesting issue concerns the limits
from existing experiments 
on the induced Lorentz-violating and time-varying couplings.
Consider again the theory \rf{prlagr}
in the supergravity background \rf{be}
with the choice $c_3=0$.
The phenomenological constraint 
$e^2(t\rightarrow\infty)\simeq4\pi/137$
implies $|A_0|\simeq 1$ and $\la\lsim 2\pi/137$.
Within this parameter space,
choose $\la = {2\pi}/{137}$ and $A_0 = \sqrt{1-\la^2}$,
which further simplifies the analysis
because it leads to a vanishing $\th$ at late times,
$\th(t\rightarrow\infty)=0$.
In fact,
the estimates below remain valid or improve
for other choices in more than 98\% 
of the allowed parameter space.

The comoving time $t$
and the time coordinate in comoving local inertial frames
agree to first order.
Assuming late times $t\gg t_0$,
we find
$e^2 \sim 2 \la \mp 8 \la^2 t_0/\sqrt{3}t$
and hence 
$\dot \al/\al \sim \pm 4 \la t_0/\sqrt{3}t^2$.
Current observational bounds on $\dot \al/\al$ at late times,
i.e., at relatively small redshifts,
are obtained from the Oklo fossil reactor
as $|\dot \al/\al| \lsim 10^{-16}$ yr$^{-1}$
\cite{oklo}.
Taking $t_n\simeq 10^{10}$ yr for the present age of the universe
then yields the estimate $t_0 \lsim 10^{6}$ yr,
consistent with the late-times assumption.

The coefficient $(k_{AF})_\mu$ for Lorentz and CPT violation
is also constrained by the Oklo data, and indeed 
constraints on axion-photon couplings 
of the form \rf{prlagr} have previously been studied 
in the context of axion and quintessence models
\cite{axph}
and CPT baryogenesis
\cite{cptb}.
In the present supergravity cosmology,
we have $\dot{N} \sim \mp 2 t_0 / \sqrt{3} \la t^2$ 
at late times,
giving $|(k_{AF})_0| \lsim 10^{-46}$ GeV.
Although model dependent,
this estimate compares favorably
with the direct observational limit 
$(k_{AF})_0 \lsim 10^{-42}$ GeV 
in Ref.\ \cite{cfj}.
Inverting the reasoning,
the latter can be used 
to bound the variation of $\al$.
We find $|\dot \al/\al| \lsim 10^{-12}$ yr$^{-1}$,
consistent with the Oklo data \cite{oklo}.

In the supergravity cosmology,
the dependence of $\al$ on time can be relatively complicated.
As an example,
the solid line in Fig.\ 1
displays the relative variation of $\al$ 
for the case $t_n/t_0 = 2000$,
as a function of the fractional look-back time $1-t/t_n$ 
to the big bang.
The parameters $\la$, $A_0$ have been changed
fractionally by parts in $10^4$
to provide an approximate match 
to the recently reported data for $\dot\al$,
also plotted in Fig.\ 1,
obtained from measurements of high-redshift spectra
over periods of approximately $0.6t_n$ to $0.8t_n$
assuming $H_0=65$ km/s/Mpc, $(\Om_m ,\Om_\La)=(0.3,0.7)$
\cite{webb}.
The parameter choices lie within the constraints on $(k_{AF})^0$,
but have no overlap with the Oklo dataset
and yield a non-asymptotic present-day value 
of the fine-structure constant. 
The solid line reflects both nonlinear features 
and a sign change for $\dot \al$.

\vskip-10pt
\begin{figure}
\centerline{\psfig{figure=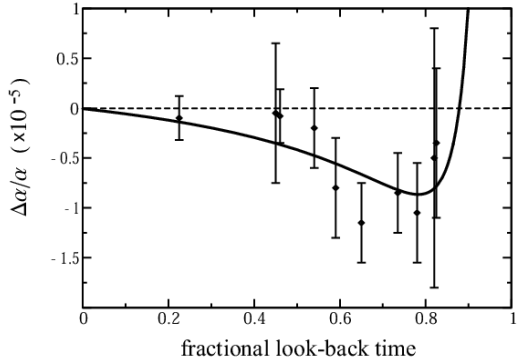,width=0.8\hsize}}
\smallskip
\caption{
Sample relative variation of the fine-structure constant
with fractional look-back time $1-t/t_n$.
}
\label{fig1}
\end{figure}

To summarize,
we have constructed an analytical supergravity cosmology 
establishing that local Lorentz and CPT violation 
can be associated with time-varying couplings.
The model shows that interesting phenomenological complications 
can appear even in an apparently simple theory 
with cosmology-induced Lorentz violation.

This work was supported in part by 
DOE grant DE-FG02-91ER40661,
by NASA grant NAG8-1770,
and by NATO grant CRG-960693.

\end{multicols}

\begin{references}
\bibitem{fn1}
The conserved symmetric energy-momentum tensor
for the lagrangian \rf{lag2}
acquires no contribution from the $N$ term 
because the latter is independent of the metric.
The other terms are positive definite.
\bibitem{fn3}
A constant timelike $(k_{AF})_\mu$ violates microcausality \cite{ck,jkk}.
The supergravity cosmology may avoid this,
but a complete analysis of this is outside our present scope.
\end{references}
\end{document}